\documentclass[12pt,preprint]{aastex}

\slugcomment{Submitted to ApJ}

\shorttitle{Kepler-15}
\shortauthors{Endl et al.}

\begin{document}

\title{The First {\it Kepler} Mission Planet Confirmed With The
Hobby-Eberly Telescope: Kepler-15b, a Hot Jupiter Enriched In Heavy Elements. 
\footnote{Based on observations obtained with the Hobby-Eberly Telescope, which is a joint project of the 
University of Texas at Austin, the Pennsylvania State University, Stanford University, 
Ludwig-Maximilians-Universit\"at M\"unchen, and Georg-August-Universit\"at G\"ottingen.}}

\author{Michael Endl, Phillip J. MacQueen, William D. Cochran}
\affil{McDonald Observatory, The University of Texas at Austin,  
    Austin, TX 78712}
\author{Erik Brugamyer}
\affil{Astronomy Department, The University of Texas at Austin,  
    Austin, TX 78712}
\author{Lars A. Buchhave}
\affil{Niels Bohr Institute, University of Copenhagen, Denmark
Centre for Star and Planet Formation, University of Copenhagen, Denmark}
\author{Jason Rowe}
\affil{SETI Institute, Moffett Field, CA 94035}
\author{Phillip Lucas}
\affil{Centre for Astrophysics Research, University of Hertfordshire, College Lane, Hatfield AL10 9AB}
\author{Howard Issacson}
\affil{Department of Astronomy, University of California, Berkeley, CA 94720}
\author{Steve Bryson, Steve B. Howell}
\affil{NASA-Ames Research Center, Moffett Field, CA 94035}
\author{Jonathan J. Fortney}
\affil{Department of Astronomy and Astrophysics, University of California, Santa Cruz, CA 95064}
\author{Terese Hansen}
\affil{Niels Bohr Institute, University of Copenhagen, Denmark}
\author{William J. Borucki} 
\affil{NASA-Ames Research Center, Moffett Field, CA 94035}
\author{Douglas Caldwell}
\affil{NASA-Ames Research Center, Moffett Field, CA 94035}
\author{Jessie L. Christiansen}
\affil{NASA-Ames Research Center/SETI Institute, Moffett Field, CA 94035}
\author{David R. Ciardi}
\affil{NASA Exoplanet Science Institute/Caltech, Pasadena, CA 91125}
\author{Brice-Olivier Demory}
\affil{Department of Earth, Atmospheric and Planetary Sciences, 
Massachusetts Institute of Technology, Cambridge, MA 02139} 
\author{Mark Everett}
\affil{NOAO, 950 N. Cherry Ave., Tucson, AZ 85719}
\author{Eric B. Ford}
\affil{Astronomy Department, University of Florida, 211 Bryant Space Sciences Center, Gainesville, FL 32111}
\author{Michael R. Haas}
\affil{NASA-Ames Research Center, Moffett Field, CA 94035}
\author{Matthew J. Holman}
\affil{Harvard-Smithsonian Center for Astrophysics, 60 Garden Street, Cambridge, MA 02138} 
\author{Elliot Horch}
\affil{Southern Connecticut State University, New Haven, CT 06515}
\author{Jon M. Jenkins}
\affil{SETI Institute/NASA-Ames Research Center, Moffett Field, CA 94035}
\author{David J. Koch} 
\affil{NASA-Ames Research Center, Moffett Field, CA 94035}
\author{Jack J. Lissauer}
\affil{NASA-Ames Research Center, Moffett Field, CA 94035}
\author{Pavel Machalek}
\affil{SETI Institute, 189 North Bernardo Ave 100, Mountain View, CA 94043}
\author{Martin Still}
\affil{NASA-Ames Research Center, Moffett Field, CA 94035}
\author{William F. Welsh}
\affil{Astronomy Department, San Diego State University, San Diego, CA 92182}
\author{Dwight T. Sanderfer}
\affil{SETI Institute/NASA Ames Research Center, Moffett Field, CA 94035}
\author{Shawn E. Seader}
\affil{SETI Institute/NASA Ames Research Center, Moffett Field, CA 94035}
\author{Jeffrey C. Smith}
\affil{SETI Institute/NASA Ames Research Center, Moffett Field, CA 94035}
\author{Susan E. Thompson}
\affil{SETI Institute/NASA Ames Research Center, Moffett Field, CA 94035}
\author{Joseph D. Twicken}
\affil{SETI Institute/NASA Ames Research Center, Moffett Field, CA 94035}

\begin{abstract}

We report the discovery of Kepler-15b, a new transiting exoplanet detected by NASA's {\it Kepler} mission. 
The transit signal with a period of 4.94 days was detected in the quarter 1 (Q1){\it Kepler} photometry.
For the first time, we have used the High-Resolution-Spectrograph (HRS) at the Hobby-Eberly Telescope (HET)
to determine the mass of a {\it Kepler} planet via precise radial velocity (RV) measurements.
The 24 HET/HRS radial velocities (RV) and 6 additional measurements
from the FIES spectrograph at the Nordic Optical Telescope (NOT) reveal a Doppler signal with the
same period and phase as the transit ephemeris. 
We used one HET/HRS spectrum of Kepler-15 taken without the iodine cell to determine accurate
stellar parameters. The host star is a metal-rich ([Fe/H]=$0.36\pm0.07$) G-type main sequence 
star with T$_{\rm eff}=5515\pm124$~K. The amplitude of the RV-orbit yields a mass of the planet of 
$0.66\pm0.1$~M$_{\rm Jup}$. The planet has a radius of $0.96\pm0.06$~R$_{\rm Jup}$ and 
a mean bulk density of $0.9\pm0.2$ g\,cm$^{-3}$. The planetary radius resides on the lower envelope for
transiting planets with similar mass and irradiation level. This suggests significant enrichment of the
planet with heavy elements. We estimate a heavy element mass of 30-40~M$_{\oplus}$ within Kepler-15b.

\end{abstract}


\section{Introduction}

The {\it Kepler} mission is designed to provide the very first estimate of the frequency of 
Earth-size planets in the habitable zone of Sun-like stars. The {\it Kepler} spacecraft 
continuously monitors $156,453$ stars (Borucki et al.~2011)
to search for the signatures of transiting planetary companions. 
The mission is described in detail in Borucki et al.~(2010). The constant 
stream of exquisite {\it Kepler} photometry makes the detection of transits of giant planets relatively easy. 

In this paper we describe the first giant planet from the {\it Kepler} mission confirmed
with the Hobby-Eberly Telescope (HET) at McDonald Observatory. Additional RV measurements were also collected with the
FIbre--fed \'Echelle Spectrograph (FIES) at the 2.5\,m Nordic Optical Telescope (NOT), 
that are fully consistent with the HET/HRS results. In the following sections we describe the {\it Kepler} photometry
for Kepler-15 and the subsequent ground-based follow-up observations to reject a false-positive and to confirm the planet.
Finally we will discuss the radius of Kepler-15b and the planet's internal composition.    

\section{{\it Kepler} photometry and transit signature} 

\begin{figure}[t]
\includegraphics[angle=0,scale=0.85]{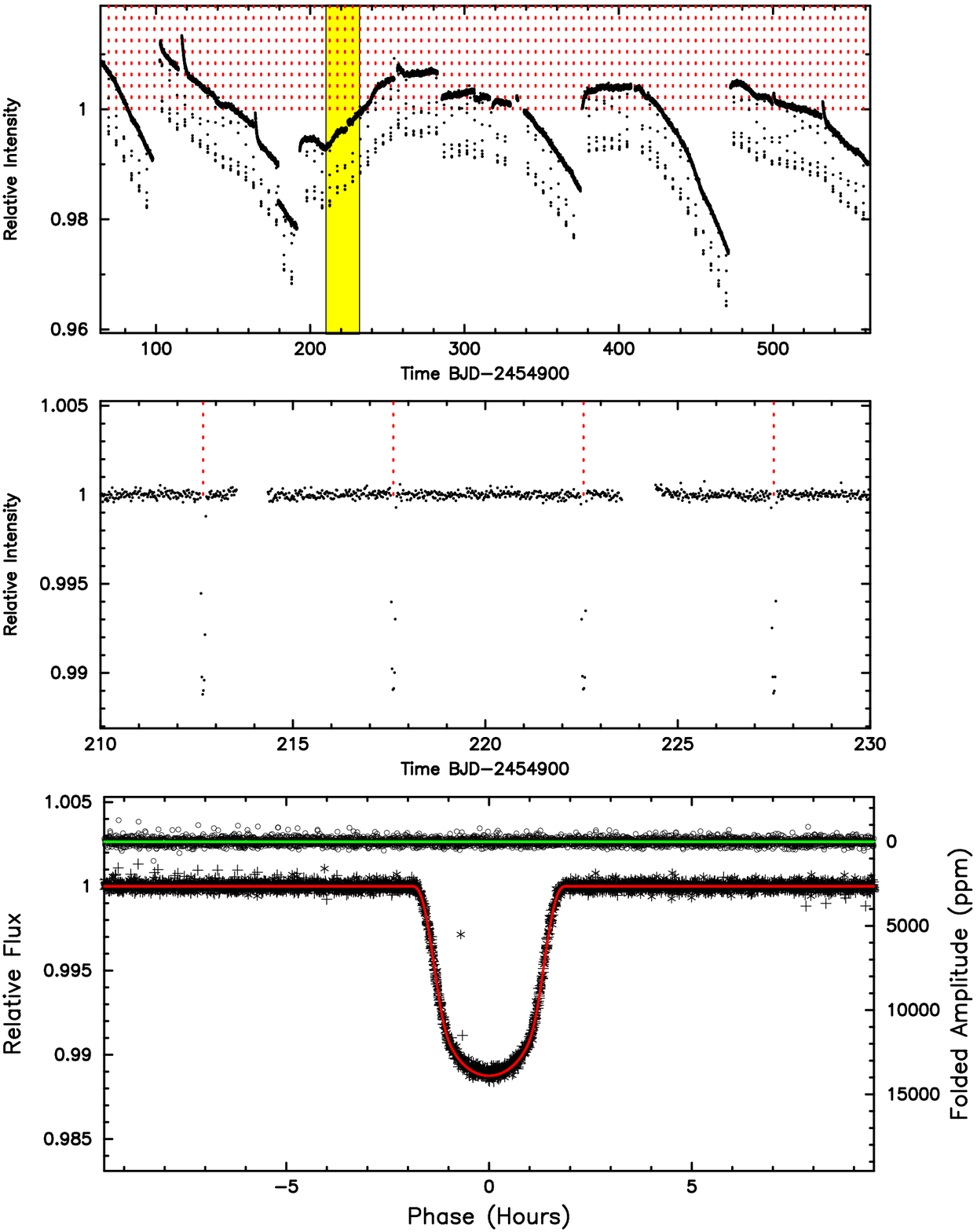}
\caption{The {\it Kepler} light curve of Kepler-15. 
The top panel displays the Q1 to Q6 time series photometry. The long-term trends are primarily caused by
focus changes in the telescope and by transitions to different detectors (due to spacecraft rotation). 
The (red) vertical dotted lines mark the location of the transits. The middle panel shows a close-up of the 
(yellow) rectangle in the upper panel (with trends removed), and the     
bottom panel contains the data phased to the transit period of 4.943\,days. 
We also show the light curve at opposite phase slightly above the
transit light curve. It shows no sign of a secondary eclipse.
\label{lc}}
\end{figure}

Kepler-15 was already identified as a planet candidate in the 35 days of the first quarter (Q1) of {\it Kepler} photometry and
was assigned the Kepler-Object-of-Interest (KOI) identifier KOI-128. The target star has a Kepler magnitude (K$_{\rm p}$) of
13.76. Processing of the photometry was carried out using the
standard {\it Kepler} pipeline (Jenkins et al.~2010a). The data were sampled at the typical 30 minute ``long cadence''.
Figure\,\ref{lc} displays the {\it Kepler} light curve for Kepler-15. 
The top panel shows the Photometric Analysis (PA) light curve for Q1 through Q6. The middle panel displays a small segment
of these data that have been processed by the Presearch Data Conditioning (PDC) software to remove spacecraft data artifacts.
The lower panel shows the data phased to the candidate period of $4.943$~days. 
The transit has a duration of 3.5~hours and a photometric depth of 12.3\, mmag (1.2$\%$).
The period of the transit is $4.9427813\pm0.000002$\,days. 

The radius of the star is estimated in the Kepler Input Catalog (KIC) as $1.4$~R$_{\odot}$. Such a large stellar radius yields a rather
large radius for the companion of $\approx 1.5$~R$_{\rm Jup}$. This large planetary radius prompted our
ground-based follow-up campaign to confirm and characterize this system. 
If such a large radius for the planet were confirmed, the planet would most-likely belong to the family of inflated hot Jupiters, 
with a very low mean density, similar to Kepler-7b (Latham et al.~2010).
    
\subsection{Difference Image Analysis}

\begin{figure}
\includegraphics[angle=0,scale=1.0]{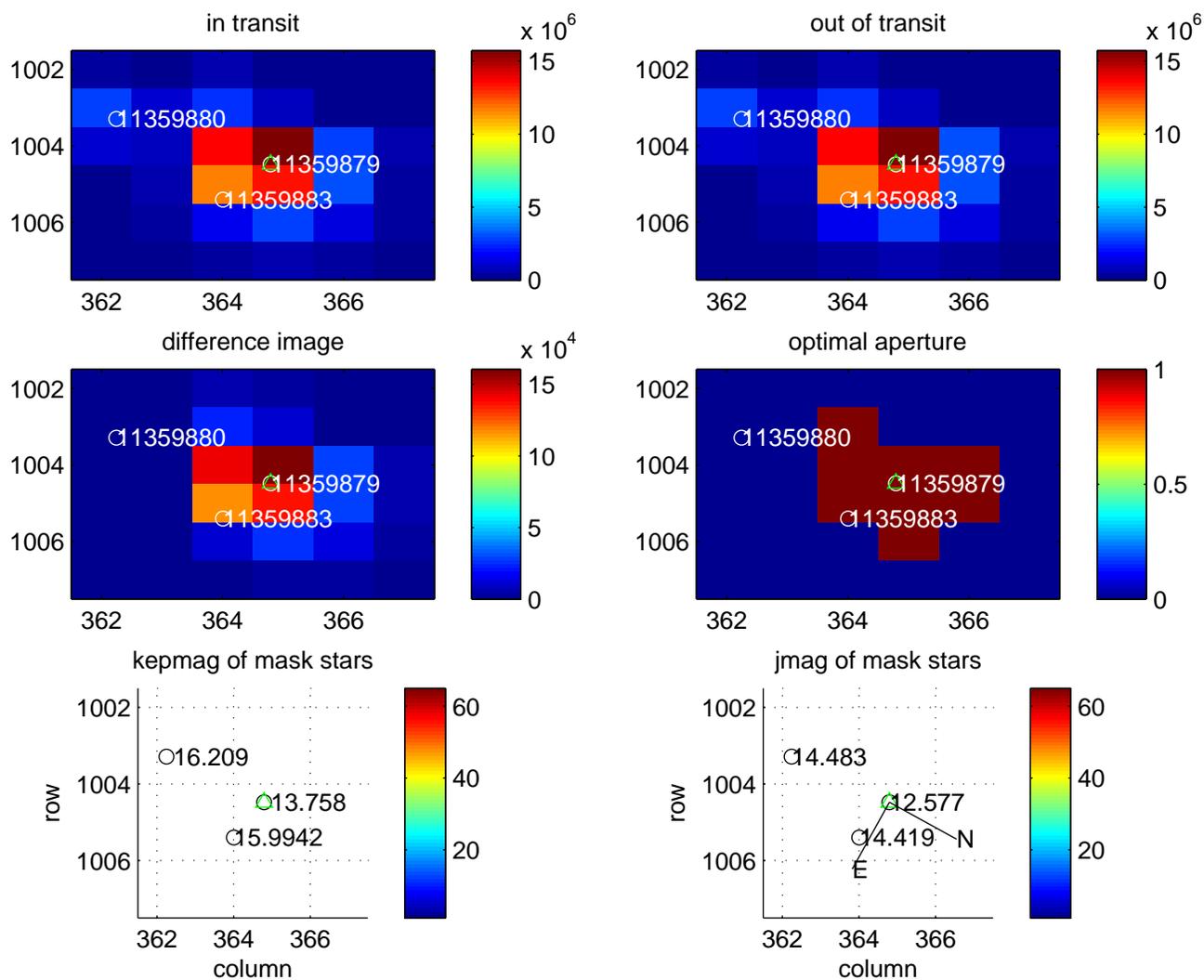}
\caption{Difference images for Kepler-15 (KIC 11359879). Each panel shows the same section of the CCD (x-axis is the pixel column and y-axis is the pixel row). 
One {\it Kepler} pixel is 3.96\,arcsec$^{2}$.
The color coding represents the flux level in each pixel. The top two panels display the in- and out of transit images and the difference image is shown in
the middle left panel. The optimal aperture for the photometry can be seen for comparison in the middle right panel. The lower two panels show the field, its orientation 
on the chip, and the K$_{\rm p}$ and J-band magnitudes of nearby KIC stars. 
\label{diff}}
\end{figure}

To eliminate the possibility that the transit signatures are due to transits on a background star, the change in centroid location during transit was 
examined using the difference image method described in Torres et al.~(2011).  This method fits the measured {\it Kepler} Pixel Response Function (PRF) 
to a difference image formed from the average 
in-transit and average out-of-transit pixel images. This difference image method has the advantage of directly measuring the location of the transiting signal.  In addition, 
the PRF is fit to the out-of-transit image to measure the position of the target star.
An example of the 
average images including the optimal aperture pixels used to create the light curve and the local stellar scene is shown in Fig\,\ref{diff}. 
The difference image completely rules out that the transits occur on the second star in the aperture (KIC 11359883). A transit on KIC 11359883 would appear 
in the difference image as a star centered on the pixel containing KIC 11359883.

\begin{figure}
\includegraphics[trim = 24mm 20mm 40mm 15mm, clip, scale=0.66]{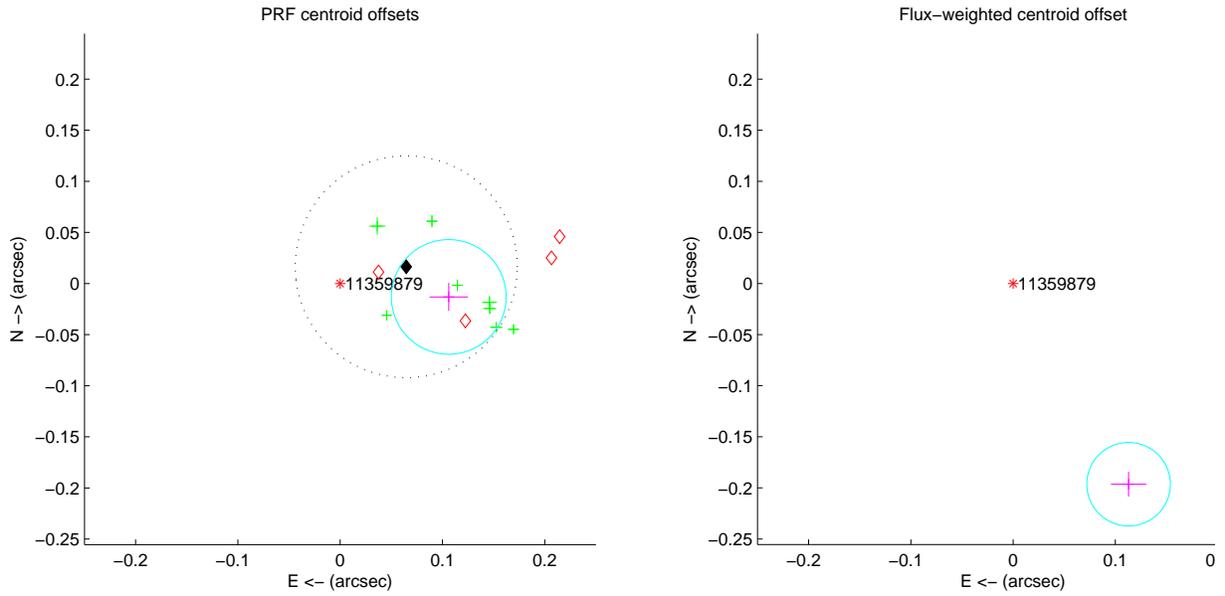}
\caption{Pixel Response Function (PRF) centroid offset (left panel) and flux-weighted centroid offset (right panel) for Kepler-15 (KIC 11359879).
The observed centroid is displayed as large cross with a solid circle showing the 3-sigma uncertainty. The left panel also contains the individually observed 
PSF centroids for Q1 through Q7 (small crosses) as well as the centroid results from our test (see text for details) to estimate the effect of crowding (small diamonds). 
A systematic offset to the west of 0.1 arcsecs due to crowding is observed and reproduced by the test results.}      
\label{centroid}
\end{figure}

These centroids 
are then subtracted to provide the offset of the transit signal location from the target star. The offsets from Q1 through Q8 are shown as the green crosses in the left 
panel of Fig.\,\ref{centroid}, where the arms of the green crosses show the uncertainty in RA and DEC.  We see that in all quarters the transit signal location 
is consistently offset to the west by about 0.1 arcsec. The robust average across quarters, weighted by the quarterly uncertainty, is shown by the magenta cross, with the 
solid circle giving its 3-sigma uncertainty radius. This average centroid observation is offset by about 0.1 arcsec with a significance of 5.7 sigma.  The right panel of 
Fig.\,\ref{centroid} shows the transit signal source estimated by correlating the transit model with observed photocenter motion (Jenkins et al.~2011b).  
Photocenter motion also shows a statistically significant (17 sigma) transit signal location offset, but the offsets from the two methods are in significant disagreement, 
suggesting that these offsets are due to measurement bias.

The uncertainty in PRF-fit centroids is based on the propagation of pixel-level uncertainty and does not include a possible PRF fit bias. Sources of PRF fit bias include 
scene crowding, because the fit is of a single PRF assuming a single star, as well as PRF error. The measured offset is the difference between the centroids of the 
difference and out-of-transit images, so common biases such as PRF error should cancel. Bias due to crowding, however, will not cancel because, to the extent that 
variations in other field stars are not correlated with transits, field stars will not contribute to the difference image. In other words the difference image will have the 
appearance of a single star where the transit occurs, so there is no crowding bias in the difference image PRF fit.

To investigate the possibility that the observed offsets are due to PRF-fit bias caused by crowding we modeled the local scene using stars from the Kepler Input Catalog 
supplemented by UKIRT observations (see section 3.2) and the measured PRF, induced the transit on Kepler-15 in the model, and performed the above PRF fit analysis on the 
model 
difference and out-of-transit images.  The resulting model offsets for quarters one through four is shown on Figure\, \ref{centroid} as open diamonds. (Only 
four quarters are shown because the model is very nearly periodic with a period of one year.) The robust average of the model offsets, again weighted by propagated 
uncertainty, is shown as the filled diamond, with the dotted circle showing the average model 3-sigma uncertainty.  We see that the model points are consistently offset to 
the west, and the observed average is well-contained within the model 3-sigma uncertainty.  
This is consistent with the observed in-transit centroid offsets being due to PRF fit bias (mostly) due to 
crowding. We therefore can be highly confident that the transit signal is due to transits on Kepler-15.

\section{Confirmation by ground-based follow-up observations}

\subsection{Reconnaissance Spectroscopy}

As part of the {\it Kepler} Follow-up Observing Program (FOP) strategy, we first obtained two reconnaissance spectra of Kepler-15 with the
Tull Coud\'e Spectrograph (Tull et al.\,1995) at the Harlan J. Smith 2.7\,m Telescope at McDonald Observatory. A comparison with a library of stellar templates yielded the
following stellar parameters: T$_{\rm eff}=5500$~K, log g = 4.0 (first spectrum), log g = 4.5 (second spectrum) and v$\sin i$ = 2~km\,s$^{-1}$.
The absolute RV of Kepler-15 is $-20$~km\,s$^{-1}$ and the two measurements differ by less than $1$~km\,s$^{-1}$ (which is within the measurement uncertainty) 
between the two visits (which were separated by one month). These results exclude the scenario where a grazing eclipsing binary produces a false-alarm (a binary would have 
produced a significant RV shift between the two reconnaissance spectra). 

\subsection{Imaging}

\begin{figure}
\includegraphics[angle=0,scale=1.0]{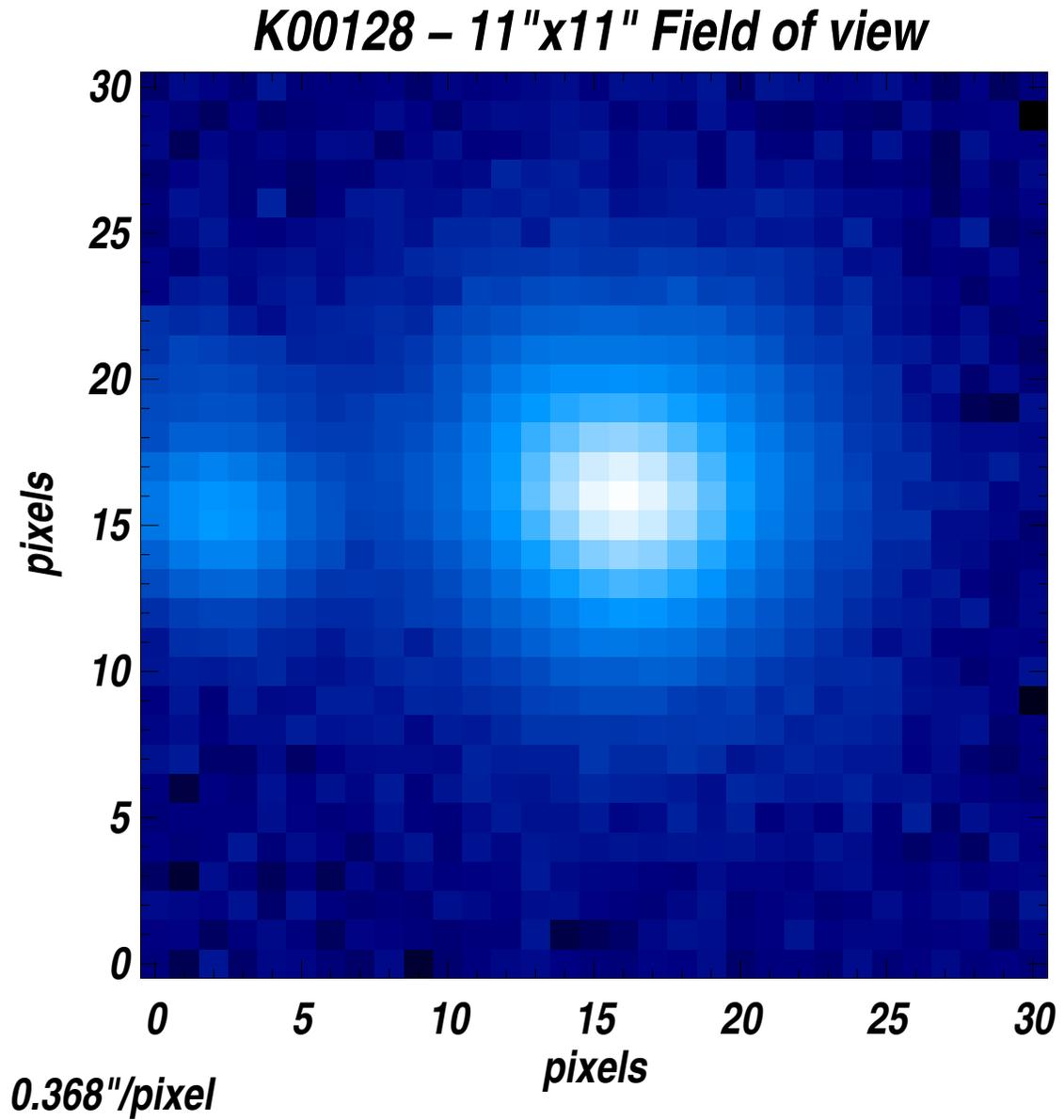}
\caption{I-band image of the field centered on Kepler-15 (KOI-128) taken with the 1\,m telescope at Lick Observatory. The scale is $11\times11$~arcseconds, north is
up and east is left. KIC 11359883, the other KIC star is the Kepler-15 aperture, is visible 4 arcseconds to the east.
\label{lick}}
\end{figure}

A seeing limited image of the field around Kepler-15 was obtained at Lick Observatory's 1\,m Nickel telescope using the Direct Imaging Camera. A single one-minute exposure
was taken in the I-band (7500-10500\,\AA), resulting in an image with seeing of approximately 1.5 arcseconds. Observations occurred under
clear skies and new moon during an observing run in 2010 July 8-10. The I-band image is shown in Figure~\ref{lick}. With the exception of the nearby star 
KIC 11359883 (K$_{\rm p}=15.99$), no other object is detected inside the optimal aperture of Kepler-15.

\begin{figure}
\includegraphics[angle=0,scale=0.6]{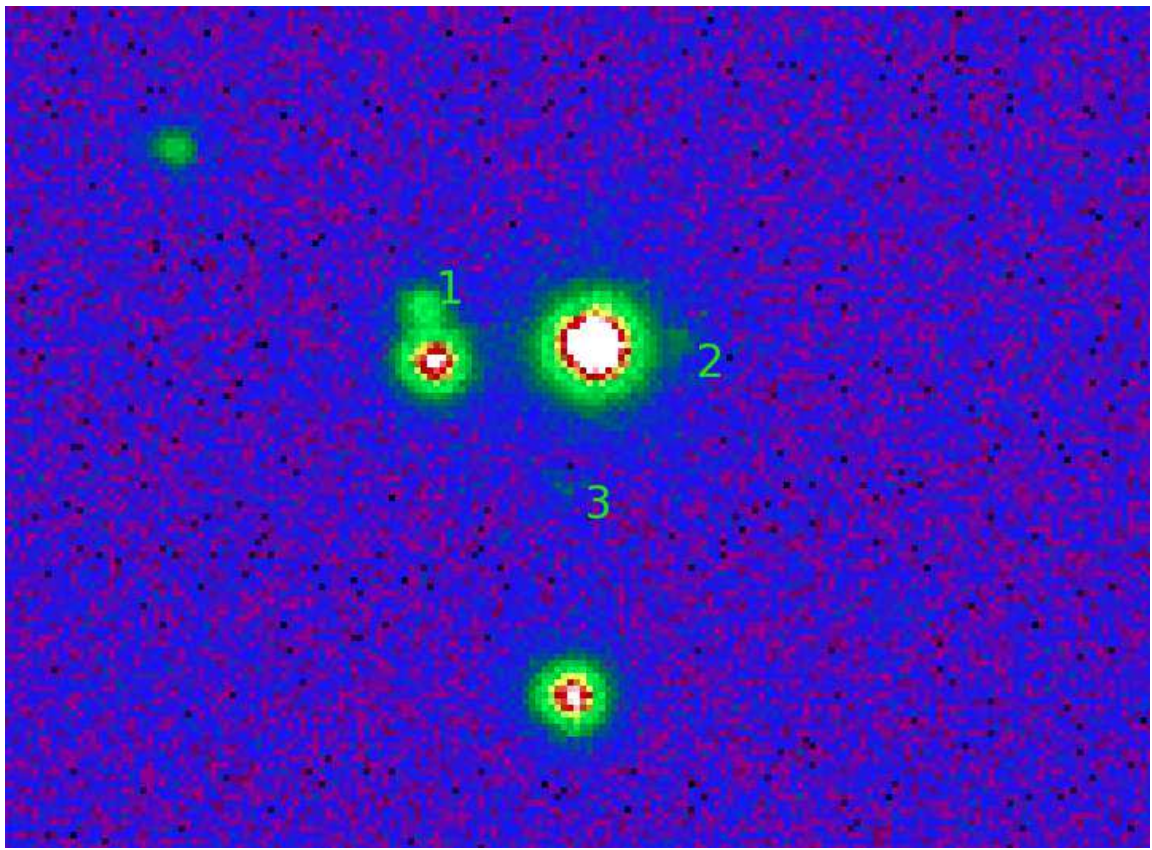}
\caption{J-band image of the field around on Kepler-15 (center) taken with UKIRT. North is up and east is left.
The scale is $\sim 25 \times 40$~arcsecs.
We detect three additional objects located within the optimal aperture of Kepler-15 (besides KIC 11359883). 
Object nr.1 and 2 are real, but nr.3 is an artifact. The two other stars seen in this image to the far east and south
are located outside the optimal aperture.
\label{ukirt}}
\end{figure}

A deep J-band image of the field was obtained at UKIRT (see Figure~\ref{ukirt}). 
Three additional objects that are located within the optimal aperture are detected in the UKIRT image. 
Object Nr.1 is a star close to KIC 11359883 and estimated to have K$_{\rm p}=19.4\pm0.6$. Object Nr.2 is very close to Kepler-15 and 
we estimate K$_{\rm p}=21.4\pm0.9$ . Object Nr. 3 is an
artifact caused by electronic cross-talk. We estimate the K$_{\rm p}$ values from the measured J-band magnitudes and the 
typical K$_{\rm p}$-J color according to the 
Besancon synthetic Galactic population model (Robin et al.\,2003) for stars of this J magnitude at this place on the sky. Object Nr.1 is 
also barely visible in the wings of the PSF of KIC 11359883 in the Lick I-band image.

We have also obtained speckle observations at the WIYN 3.5-m telescope located on Kitt Peak.
The observations make use of the Differential Speckle Survey Instrument (DSSI), a
recently upgraded speckle camera described in Horch et al. (2010) and Howell et al.,
(2011). The DSSI provides simultaneous observations in two filters by employing a
dichroic beam splitter and two identical EMCCDs as the imagers. We observed Kepler-15 simultaneously in "V" and
"R" bandpasses where "V" has a central wavelength of 5620\,\AA, and "R" has a central
wavelength of 6920\,\AA, and each filter has a FWHM=400\,\AA.
The details of how we obtain, reduce, and analyze the speckle results and specifics about how they are used to eliminate
false positives and aid in transit detection are described in Howell et al.~(2011).

\begin{figure}
\includegraphics[angle=0,scale=0.6]{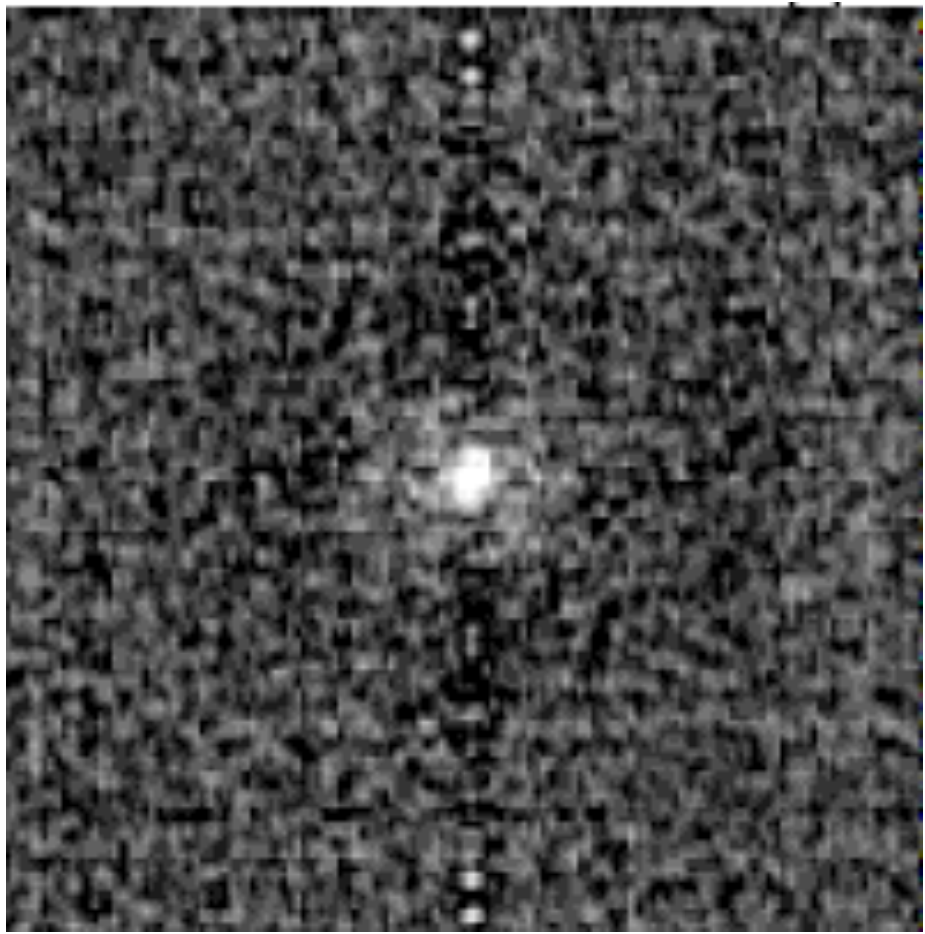}
\caption{
Reconstructed R-band speckle image of Kepler-15 taken with the WIYN telescope at Kitt Peak.
No additional stars are detected within the annulus from 0.05 to 1.8 arcsec to a limit
of (5$\sigma$) 3.52 magnitudes fainter than the target star. The image is $2.8\times2.8$~arcsecs.
\label{speckle}}
\end{figure}

The speckle observations of the Kepler-15
were obtained on 2010 October 24 (UT) and consisted of five sets of 1000, 40 msec
individual speckle images. Our R-band reconstructed image is
shown in Figure\,\ref{speckle} with details of the image composition described in Howell et al.~(2011).
Along with a nearly identical V-band reconstructed image, the speckle results reveal
no companion star near Kepler-15 within the annulus from 0.05 to 1.8 arcsec to a limit
of (5$\sigma$) 3.52 magnitudes fainter in R and 3.16 magnitudes fainter in V relative to the K$_{\rm p}=13.76$ target star.

As a result of the direct imaging of the field around Kepler-15 we found that two additional stars (besides KIC 11359883) are located within the optimal aperture 
of Kepler-15. However, both stars
are fainter than K$_{\rm p}=19$ and have a negligible effect on the photometry. Only KIC 11359883 (K$_{\rm p}=15.99$) has a significant effect and
we take the diluting effect of its light contribution into account for the light curve modeling. 

\subsection{Precise Radial Velocity Measurements}

We performed precise RV follow-up observations of Kepler-15 with the HET (Ramsey et al.~1998) and its HRS spectrograph (Tull~1998). 
The queue-scheduled observing mode of the HET usually leads to
the situation that on a given night, data for many different projects and with different instruments are obtained. The
observations are ranked according to priorities distributed by the HET time-allocation-committees as well as additional timing 
constraints. We entered Kepler-15 into the HET queue to be observed in a quasi-random fashion with a cadence of a few days 
to allow proper sampling of the suspected 4.9~d RV orbit.
We observed this target from 2010 March 29 until 2010 November 9. 
We collected 24 HRS spectra with the I$_2$-cell in the light path for precise RV measurements. Furthermore,
we obtained one spectrum without the I$_2$-cell to serve as a stellar ``template'' for the RV computation and to better
characterize the properties of the host star. 

Because of the faintness of this star, the HRS setup we employed for the RV observations is slightly different from 
our standard planet search RV reduction pipeline (described in detail 
in Cochran et al.~2004). We used the 2 arcsec fiber to feed the light into the HRS. The cross-disperser setting was ``600g5822'', which corresponds
to a wavelength coverage from 4814 to 6793\,\AA, thus covering the entire I$_2$ spectral range of 5000-6400\,\AA.  
We also used a wider slit to gain a higher throughput for this faint target, reducing the spectral resolving
power to $R =\lambda/\Delta\lambda = 30,\!000$ (instead of our nominal $R=60,\!000$). Moreover, two sky fibers allow us to simultaneously record the sky 
background and to properly subtract it from our data. The CCD was binned 2x2, which yields 4 pixels per resolution element. 
This new setup is better suited for observations of the faint Kepler targets. The exposure time for each observation 
was $1200$ seconds. The mean S/N-ratio of the 24 spectra is $42\pm6$ per resolution element.
As higher spectral resolution is advantageous for the template spectrum, we obtained this spectrum with $R=60,\!000$ and a 
longer exposure time of 2700 seconds. We computed precise differential RVs with our {\it Austral} I$_2$-cell data modeling algorithm (Endl et al.~2000). 
The HET/HRS RV data are listed in table\,\ref{rvstab}. The data have an overall rms-scatter of $60~{\rm m\,s}^{-1}$ and average internal errors of 
$25\pm8~{\rm m\,s}^{-1}$. 

\begin{figure}
\includegraphics[angle=270,scale=0.5]{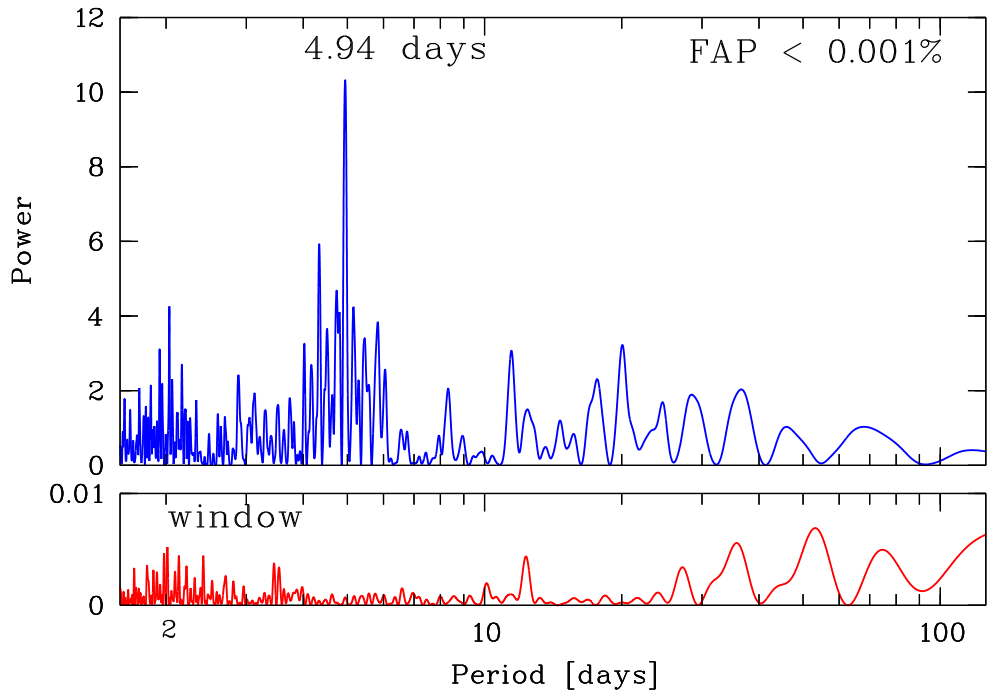}
\caption{Lomb-Scargle periodogram of the HET/HRS RV results for Kepler-15. The top panel shows the power spectrum of the periodogram with a highly
significant peak (false-alarm probability FAP $<10^{-5}$) at the transit period of 4.94 days. 
The lower panel displays the window function of the HET observations.
\label{per}}
\end{figure}

We performed a period search in our HET RV data set using the classic Lomb-Scargle periodogram (Lomb~1976, Scargle~1982). 
Fig.\,\ref{per} displays the power
spectrum over the period range from 2 to 100 days. The highest peak is located at a period of $4.94$\, days. This is an independent
confirmation of the transit period. The signal is also statistically highly significant, we estimate a false-alarm-probability (FAP) of
less than $10^{-5}$ using a bootstrap randomization scheme (K\"urster et al.~1997).

\begin{figure}
\includegraphics[angle=270,scale=0.5]{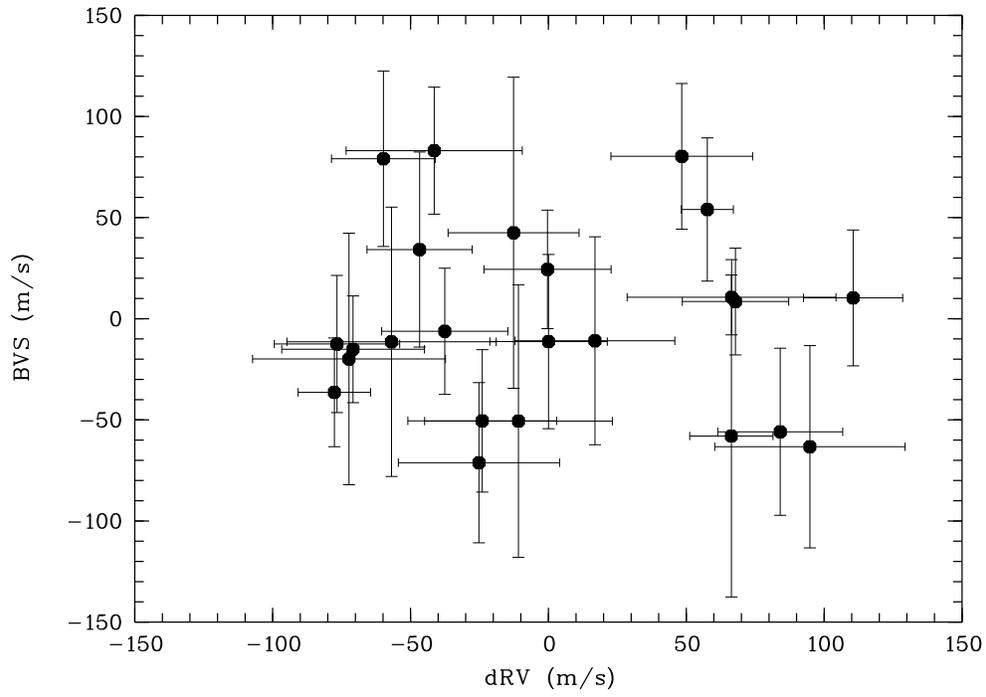}
\caption{Correlation between the bisector velocity span (BVS) and the RV measurements of the 24 HET spectra. No correlation is 
detected (linear correlation coefficient is $-0.076$). 
\label{bvs}}
\end{figure}

We have also determined line bisectors from the HET spectra. As we could use only the small fraction of the available spectral range that lies
outside the I$_2$ region ($5000-6400$\,\AA) the uncertainties in the bisector velocity span (BVS) are quite large, the average error of the
BVS measurements is $43\pm17$\,m\,s$^{-1}$ and they have a total rms-scatter of 46\,m\,s$^{-1}$. The bisector measurements are given in
table\,\ref{bisector}.
Figure\,\ref{bvs} shows the correlation plot of BVS values versus RV measurements. The linear correlation coefficient is $-0.076$ corresponds 
to a 72\% probability that the null-hypothesis of zero correlation is true. This further strengthens the case that the RV modulation is due to an
orbiting companion.

We have also taken 6 spectra between 2010 July and August using the FIbre--fed
\'Echelle Spectrograph (FIES) at the 2.5\,m Nordic Optical Telescope (NOT)
at La Palma, Spain (Djupvik \& Andersen\,2010). We used the medium and the
high--resolution fibers ($1\farcs3$ projected diameter) with resolving
powers of $R \approx 46,\!000$ and $67,\!000$,
respectively, giving a wavelength coverage of $\sim 3600-7400$\,\AA.
We used the wavelength range from approximately $\sim 4100-5600$\,\AA to determine the RVs following 
the procedures described in Buchhave et al.~(2010).
The exposure time was between 2400 and 3600 seconds, yielding a S/N from 22 to 30 
per pixel in the wavelength range used. The FIES RV results are
also given in Table\,\ref{rvstab}. 

\begin{figure}
\includegraphics[angle=270,scale=0.5]{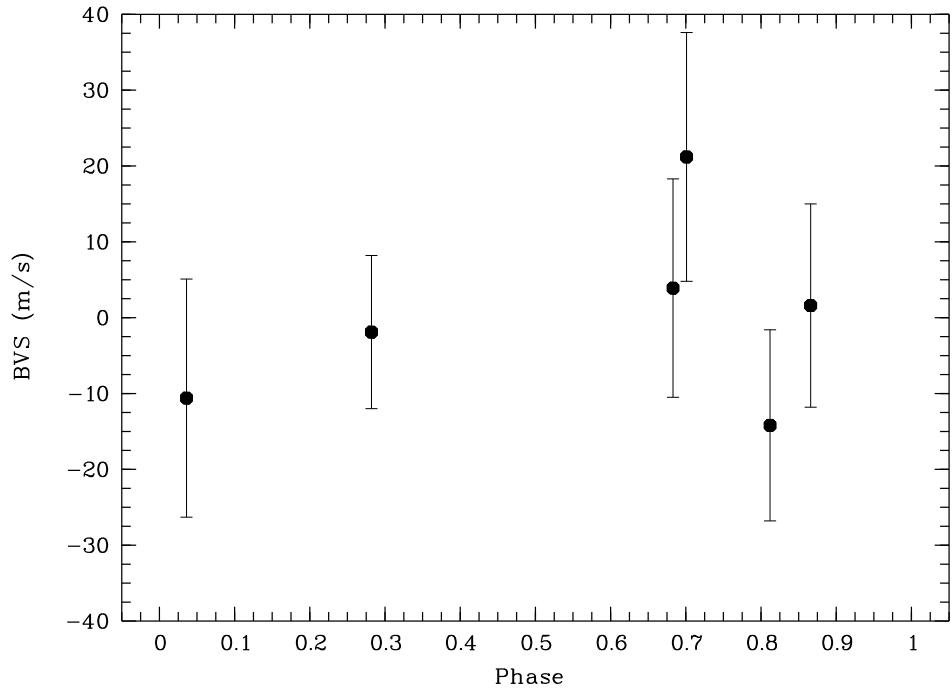}
\caption{FIES line bisectors and their uncertainties as a function of transit/orbital phase of Kepler-15. We 
detect no significant variability.
\label{bvs2}}
\end{figure}

We also determined the line bisectors for the 6 FIES spectra. 
They have a higher precision than the HRS results since
we could use the entire spectrum for the analysis (FIES does not use an I$_2$-cell). The FIES bisector
data are listed in table\,\ref{bisector} and shown as a function of orbital phase in Figure\,\ref{bvs2}.
The average uncertainty of the FIES bisector measurements is $13.8\pm2.3$\,m\,s$^{-1}$ and their total scatter
is $12.5$\,m\,s$^{-1}$. They appear to be constant within the measurement uncertainties.

\section{Results}

\subsection{Host star characterization}

We determined stellar parameters using the local thermodynamic equilibrium (LTE) line analysis and spectral synthesis code MOOG\footnote{available at
http://www.as.utexas.edu/~chris/moog.html} (Sneden 1973), together with a grid of Kurucz (1993) ATLAS9 model atmospheres.  The method used is virtually identical to that
described in Brugamyer et al.\,(2011).  
To check this method, we first measured the equivalent widths of a carefully selected list of 48 neutral iron lines and 11 singly-ionized iron lines in a
spectrum of the daytime sky, taken using the same instrumental setup and configuration as that used for Kepler-15.  MOOG force-fits abundances to match these measured
equivalent widths, using declared atomic line parameters. By assuming excitation equilibrium, we constrained the stellar temperature by eliminating any trends with
excitation potential; assuming ionization equilibrium, we constrained the stellar surface gravity by forcing the derived iron abundance using neutral lines to match that
of singly-ionized lines.  The microturbulent velocity was constrained by eliminating any trend with reduced equivalent width (=EW/$\lambda$).  Our derived stellar parameters
for the Sun (using our daytime sky spectrum) are as follows: T$_{\rm eff}$ = $5755 \pm 70$\,K, log g = $4.48 \pm 0.09$ dex, Vmic = $1.07 \pm 0.06$ km\,s$^{-1}$, 
and log {$\epsilon$} (Fe) = $7.53 \pm 0.05$ dex.

The process described above was repeated for the HET/HRS spectrum taken without the I$_2$ cell of Kepler-15.
We took the difference, on a line-by-line basis, of the derived iron abundance from each line.  Our quoted iron
abundance is therefore differential with respect to the Sun.  To estimate the rotational velocity of the star, we synthesized three 5-\AA-wide spectral regions in the
range 5640 - 5690\,\AA, and adjusted the gaussian and rotational broadening parameters until the best fit (by eye) was found to the observed spectrum.  The results of
our analysis yield the following stellar parameters for Kepler-15: T$_{\rm eff}$ = $5595\pm 120$ K, log g $= 4.23\pm 0.2$, Vmic = $1.09 \pm 0.1$ km/s, [Fe/H] = $+0.36\pm 
0.07$, and Vrot = $2 \pm 2$ km\,s$^{-1}$. The preliminary results from the reconnaissance spectroscopy (T$_{\rm eff} \approx 5500$ K, log g $= 4.0$ \& $4.5$ and Vrot = $2$ 
km\,s$^{-1}$) compare very well with this improved spectroscopic analysis.

\subsection{Orbital Solution and Planet Parameters}

\begin{figure}
\includegraphics[angle=0,scale=0.8]{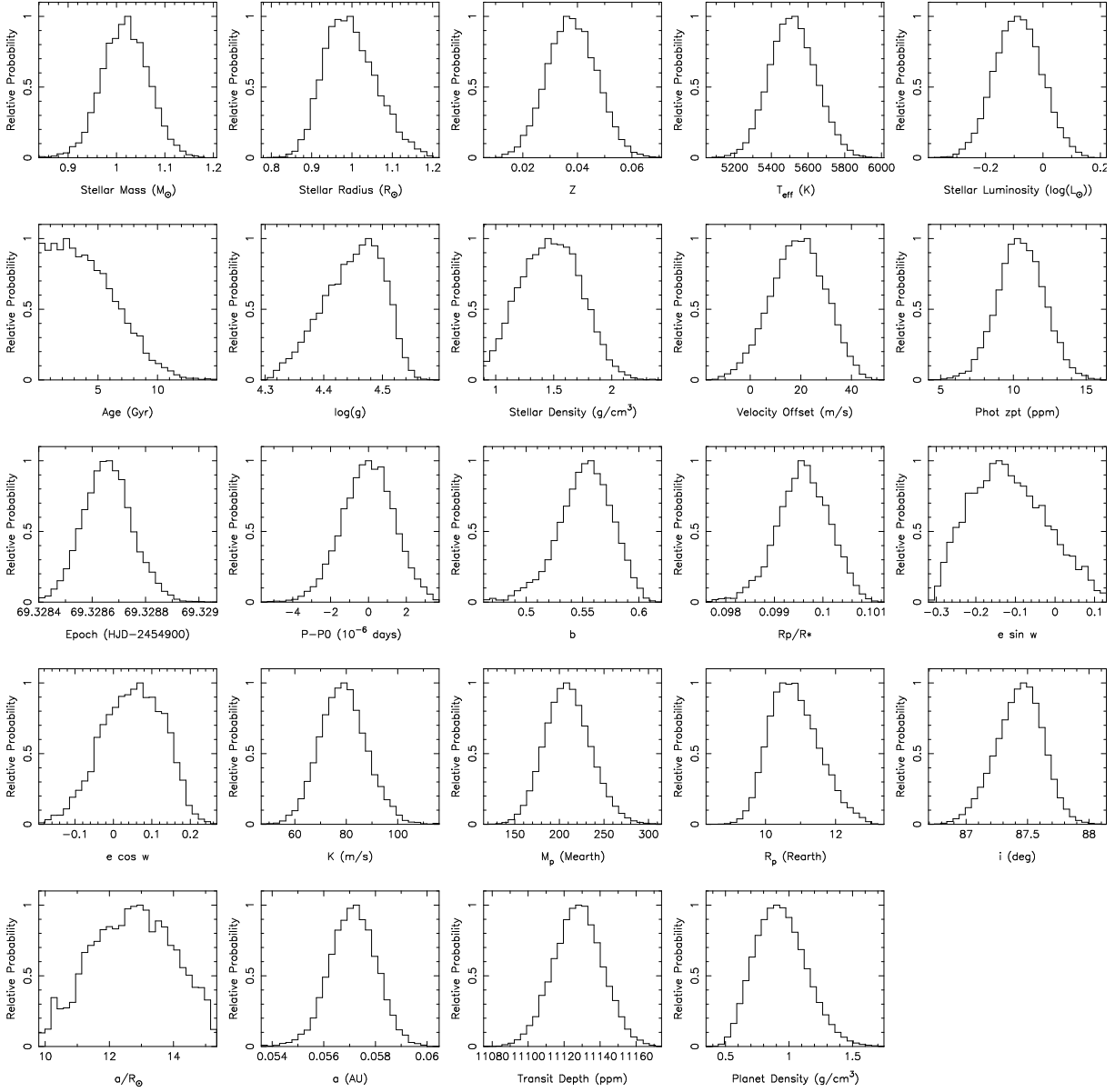}
\caption{Markov chain Monte Carlo distributions for the model parameters for the Kepler-15 system. 
The resulting values and uncertainties of the system parameters are listed in Table\,\ref{tab:planet}.
For the period value we plot the difference to P0=4.942782~days.
\label{mcmc}}
\end{figure}

We used a Markov Chain Monte Carlo (MCMC) algorithm to perform a simultaneous fit to the light curve and the 
RV results. This analysis was performed using the 24 HET RV measurements and the Q1 through Q6 {\it Kepler} photometry. 
The model fits for $\rho_{\star}$, T0, Period, b, r/R$_{\star}$, $e \sin \omega$, $e \cos \omega$, $\gamma$, and the photometric zeropoint.  
The transit shape is characterized by the 
Mandel-Agol analytic derivations (Mandel \& Algol~2002) and the planetary orbit is assumed to be Keplerian.
The best fit model is computed by simultaneously fitting radial velocity measurements and Kepler photometry and then 
minimizing the chi-square statistic with a Levenberg-Marquart method.  
To obtain probability distributions, the best fit model is used to seed a MCMC computation. 
Our MCMC algorithm employs a hybrid sampler based on Gregory (2011) that uses a 
Gibb-sampler or a buffer of previously computed chain points to generate proposals to jump to new locations in parameter space. The addition of the buffer 
allows for a calculation of vectorized jumps that allow for efficient sampling of highly correlated parameter space. The MCMC distributions are shown in
Figure\,\ref{mcmc}.

\begin{figure}
\includegraphics[angle=270,scale=0.65]{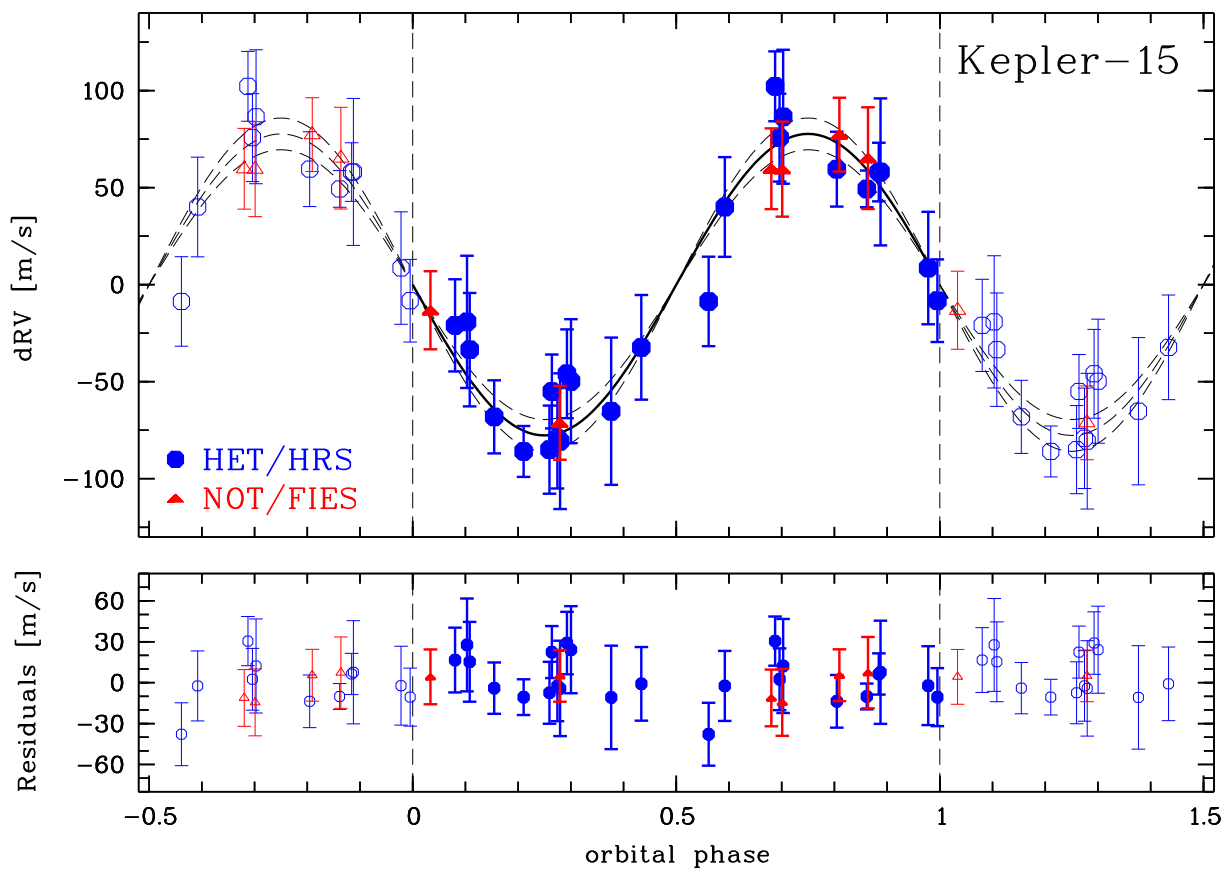}
\caption{HET/HRS (circles) and NOT/FIES (triangles) RV data and the best-fit orbital solution (solid line) phased to the transit period of
4.94\,days (top panel). The semi-amplitude $K$ is $78.7\pm9.1$\,m\,s$^{-1}$ corresponding to a mass of
$0.66\pm0.08$\,M$_{\rm Jup}$ for the planetary companion. The bottom panel shows the residuals: the 24 HRS points have a residual
rms scatter of 16.9\,m\,s$^{-1}$ and the 6 FIES points have a residual rms scatter of 9.6\,m\,s$^{-1}$. (The data are repeated for a second
cycle.)
\label{orbit}}
\end{figure}

The results of this MCMC modeling are summarized in Table\,\ref{tab:planet}. For all parameters we list the median values along with their 68\% uncertainty interval
($\pm 1\sigma$) based on the MCMC distributions.  
The transit ephemeris is $T0=69.328651^{+0.000084}_{-0.000096}$ (BJD-2454900) and the
period is $P=4.942782 \pm 0.0000013$ days. The transit has a depth of $11127.7^{+12.8}_{-14.4}$~ppm.
Taking into account the diluting effect of the other star in the aperture ($97.0\pm0.003$\% of the light in the aperture comes from
Kepler-15) we find a radius ratio of $R_{\rm planet} / R_{*} = 0.09960^{+0.00055}_{-0.00053}$. The RV semi-amplitude $K$ is $78.7^{+8.5}_{-9.1}$\,m\,s$^{-1}$. 
The orbital eccentricity was allowed as a free parameter during the modeling process, but we find
no strong indication for an eccentric orbit ($e \sin \omega=-0.123^{+0.089}_{-0.110}, e \cos \omega=0.053^{+0.086}_{-0.079}$).  
Figure\, \ref{orbit} displays the HET/HRS and the NOT/FIES results compared to the RV orbit (assuming $e=0$). The reduced $\chi^{2}$ of this fit is 0.52, indicating that
our RV error bars are slightly overestimated. The residual rms-scatter for the two RV data set is: 16.9\,m\,s$^{-1}$ for the HET/HRS data and 9.6\,m\,s$^{-1}$ 
for the NOT/FIES data.

The distribution of $\rho_{\star}$ from the model fit above and T$_{\rm eff}$ and [Fe/H] from the spectroscopic determination are used together to match 
Yonsei-Yale (Y$^{2}$) models (Yi et al.~2001). This is known as the ``$\rho_{\star}$ method'' (see e.g. Sozetti et al.~2007, Brown~2010).  
The probability distributions of the matching stellar parameters, M$_{\star}$, R$_{\star}$, age, luminosity are also 
shown in Figure\,\ref{mcmc}. 
From isochrone fitting we derive a mass for the star of $1.018^{+0.044}_{-0.052}$\,M$_{\odot}$, a radius of $0.992^{+0.058}_{-0.070}$\,R$_{\odot}$ and an age of $3.7^{+3.5}_{-1.6}$\,Gyrs.
The log \,g from the isochrone fit is $4.46^{+0.053}_{-0.050}$, 0.2 dex higher than the spectroscopically derived log\,g value of $4.23\pm 0.2$. 
A systematic difference in log\,g values for high metallicity stars has been discussed previously for Kepler-6 (Dunham et al.\,2010). We note that the difference in our case
is the opposite (the spectroscopic log\,g is lower, for Kepler-6 it was higher by 0.35 dex). However, the error bars of both methods still overlap for Kepler-15.   
All stellar parameters from the isochrone fit are also listed in Table\,\ref{tab:planet}.

The planetary orbit and parameters are derived by using the Markov chains from the model and isochrone fits. Random chains are selected from each fit to calculate the 
planet mass, radius, semi-major axis, a/R$_{\star}$ and presented in Figure\,\ref{mcmc} and Table\,\ref{tab:planet}.

\section{Discussion}

After passing all tests and observational diagnostics, from photometric centroid shifts to spectroscopic line bisectors, and 
the fact that an RV orbit in period and phase with the transit ephemeris is detected, we conclude that Kepler-15b is indeed
a new transiting planet.
All our currently available data suggest a planet with a mass of $0.66$~M$_{\rm Jup}$, a radius of 0.96~R$_{\rm Jup}$ and a mean
density of $0.9\pm0.2$\,g\,cm$^{-3}$ orbiting a metal-rich ([Fe/H]=0.36) G-type star every 4.94\,days. Kepler-15 is tied with Kepler-6 as the
most metal-rich host star of all currently published {\it Kepler} planets.      

\begin{figure}
\includegraphics[angle=0,scale=1.0]{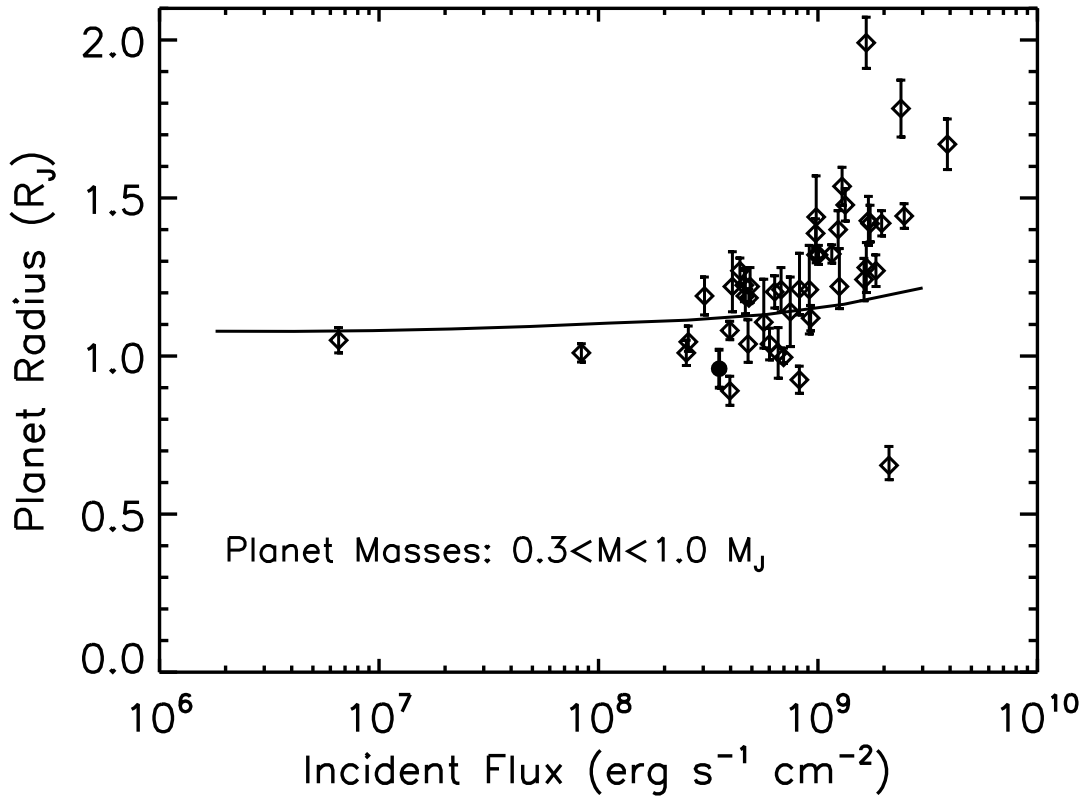}
\caption{
Planet mass vs.~radius for a collection of transiting planets with masses from 0.3 to 1.0 M$_{\rm Jup}$. Values are taken from: http://www.inscience.ch/transits/.  
Kepler-15 is shown as a filled circle.  The solid curve is for a 1~M$_{\rm Jup}$ model planet at 4.5 Gyr (Miller, Fortney, and Jackson, 2009).  
While many planets are inflated relative to this curve, Kepler-15 is clearly below it.  
While a radius-inflation mechanism could still be at work in this planet, the small radius indicates that the planet is rich in heavy elements. 
\label{radius}}
\end{figure}

Although we initially suspected a large planetary radius, our results demonstrate rather the opposite.
Figure\,\ref{radius} shows that the planet's radius is modestly smaller than planets of similar mass and irradiation level.  This suggests that the planet is more enriched in heavy elements 
than most other transiting planets. Given the high stellar metallicity, and the connection between stellar metallicity and planetary heavy elements 
(Guillot et al.~2006, Burrows et al.~2007, Miller \& Fortney~2011), this is is not surprising.  Using the tables of Fortney et al.\,(2007), we estimate a heavy element 
mass within the planet of at least 30-40 Earth masses. 

\section{Summary \& Outlook}

We report the discovery of Kepler-15b, a hot Jupiter ($P=4.94$\,d) that is enriched in heavy elements ($\approx 20\%$ in mass) in orbit around a metal-rich G-type star with
K$_{\rm p}$=13.76. 
Kepler-15b is the first giant planet from the {\it Kepler} mission that we confirmed with the HET, demonstrating the capability of this facility as
an integral part of the ground-based spectroscopic follow-up effort of the {\it Kepler} mission. In 2010 we spent a total of 65 hours on the {\it Kepler} field and collected
data for 11 {\it Kepler} candidates. Besides the planet presented here, we have confirmed several other giant planets around {\it Kepler} stars, e.g.
Kepler-17b (D\'esert et al.~2011). The remaining HET planet confirmations will be included in
a catalog of {\it Kepler} giant planets (Caldwell et al. in prep.). 

The HET will obtain a major upgrade to its secondary tracker assembly to allow a wider field of view. This upgrade will start in fall 2011. During the telescope downtime the
HRS will also undergo a major upgrade, including more efficient optics and fibers, as well as image slicers, to boost the overall throughput 
by several factors. Once the HET is
back on sky in early 2012, these improvements should allow us to use the HET/HRS also for the confirmation and validation of low mass 
{\it Kepler} planet candidates in the Neptune and Super-Earth 
range, similar to Kepler-10b (Batalha et al.\,2011).

\acknowledgments
Funding for this Discovery mission is provided by NASA's Science Mission Directorate.
The Hobby-Eberly Telescope (HET) is a joint project of the University of
Texas at Austin, the Pennsylvania State University, Stanford University,
Ludwig-Maximilians-Universit\"{a}t M\"{u}nchen,
and Georg-August-Universit\"{a}t G\"{o}ttingen.
The HET is named in honor of its principal benefactors,
William P. Hobby and Robert E. Eberly. 
Based in part on observations made with the Nordic Optical
Telescope, operated on the island of La Palma jointly by
Denmark, Finland, Iceland, Norway, and Sweden, in the Spanish
Observatorio del Roque de los Muchachos of the Instituto de
Astrofisica de Canarias.


\begin{deluxetable}{lrrr}
\tablecolumns{4}
\tablewidth{0pt}
\tablecaption{Radial velocity measurements for Kepler-15
\label{rvstab}}
\tablehead{
\colhead{BJD [d]} & {dRV[m\,s$^{-1}$]} & {err[m\,s$^{-1}$]} & {Telescope}
}
\startdata
2455284.981307  &    57.6  &  9.5 & HET \\
2455286.970962  &   -46.7  &  19.1 & HET \\
2455335.858420  &  -59.8   &  18.8 & HET \\
2455337.868951  &    -0.3 &  23.1 & HET \\
2455346.841507    & -56.9 &  38.0 & HET \\
2455349.813386   &   16.9 &  29.0 & HET \\
2455357.791818   &   48.4 &  25.7 & HET \\
2455359.784683  &     0.1 &  21.3 & HET \\
2455363.784770  &    67.8 &  19.3 & HET \\
2455370.733856   &  -77.6 &  13.1 & HET \\
2455395.690132  &   -76.7  & 22.7 & HET \\
2455397.881162   &   94.8 &  34.5 & HET \\
2455399.884842   &  -25.2 &  29.2 & HET \\
2455405.675158   &  -72.3  & 35.0 & HET \\
2455470.693433  &   -24.0 &  27.0 & HET \\
2455494.620347  &   -70.9 &  25.9 & HET \\
2455497.633684 &     66.4 &  15.1 & HET \\
2455498.603079  &   -12.6 &  23.8 & HET \\
2455501.603733 &    110.5 &  18.0 & HET \\
2455502.592532  &    66.4 &  37.9 & HET \\
2455504.595288  &   -37.6 &  22.9 & HET \\
2455506.589121  &    84.1 &  22.7 & HET \\
2455508.600869  &   -10.9 &  34.0 & HET \\
2455509.575509  &   -41.4 &  31.9 & HET \\
\hline
2455378.649338 & 148.5  &     19.0 & NOT \\
2455384.700553 & 58.0   &    20.1 & NOT \\ 
2455423.403627 & 136.4  &     26.2 & NOT \\
2455425.457207 & 0.0    &   19.0 & NOT \\
2455427.440796 & 130.9  &     20.8 & NOT \\
2455432.470155 & 130.7  &     24.5 & NOT \\
\hline
\enddata
\end{deluxetable}

\begin{deluxetable}{lrrr}
\tablecolumns{4}
\tablewidth{0pt}
\tablecaption{Bisector measurements for Kepler-15
\label{bisector}}
\tablehead{
\colhead{BJD [d]} & {bisector[m\,s$^{-1}$]} & {bserr[m\,s$^{-1}$]} & {Spectrograph}
}
\startdata
2455284.981307  & 54.0  & 35.4  & HRS \\
2455286.970962  & 34.2  & 48.2  & HRS \\
2455335.858420  & 79.1 & 43.4  & HRS \\
2455337.868951  & 24.4  & 29.3  & HRS \\
2455346.841507  & -11.4   & 66.6  & HRS \\
2455349.813386  & -10.9  & 51.4  & HRS \\
2455357.791818  &  80.3  & 36.0  & HRS \\
2455359.784683  & -11.3 & 43.1  & HRS \\
2455363.784770  &  8.5 & 26.4  & HRS \\
2455370.733856  & -36.4   & 26.9  & HRS \\
2455395.690132  &  -12.5 &  33.9 & HRS \\
2455397.881162  & -63.3  & 50.0  & HRS \\
2455399.884842  & -71.2  & 39.6  & HRS \\
2455405.675158  &  -19.9 & 62.2  & HRS \\
2455470.693433  &  -50.5 & 35.2  & HRS \\
2455494.620347  & -15.1  & 26.4  & HRS \\
2455497.633684  &  -58.0 & 79.6  & HRS \\
2455498.603079  &  42.5 & 76.9  & HRS \\
2455501.603733  &  10.3 & 33.6  & HRS \\
2455502.592532  &  10.6 & 18.6  & HRS \\
2455504.595288  &   -6.2  & 31.2  & HRS \\
2455506.589121  &  -55.9 & 41.3  & HRS \\
2455508.600869  &  -50.6 & 67.4  & HRS \\
2455509.575509  &  83.1 & 31.4  & HRS \\
\hline
2455378.649338 & -14.2 & 12.6 & FIES \\
2455384.700553 & -10.6 & 15.7 & FIES \\
2455423.403627 & 1.6 &  13.4  & FIES \\
2455425.457207 & -1.9 & 10.1  & FIES \\
2455427.440796 & 3.9  & 14.4  & FIES \\
2455432.470155 & 21.2 & 16.4  & FIES \\
\hline
\enddata
\end{deluxetable}

\begin{deluxetable}{lrrrr}
\tablecolumns{3}
\tablewidth{0pt}
\tablecaption{Parameters of the Kepler-15 transiting system
\label{tab:planet}}
\tablehead{
\colhead{Parameter [unit]} & {value} & {+1$\sigma$} & {-1$\sigma$} & {notes}
}
\startdata
KIC & 11359879 & & & \\
KOI & 128 & & & \\
K$_{\rm p}$[mag] & 13.76 & & & \\
RV [km\,s$^{-1}$] & -20.0 & +1.0 & -1.0 & \\
\hline
Period [days]& 4.942782 & +0.0000013 & -0.0000013 & \\
T0 [BJD] & 2454969.328651 & +0.000084 & -0.000096 & \\
$\rho_{\star}$ [g\,cm$^{-3}$] & 1.47 & +0.26 & -0.28 & \\
b & 0.554 & +0.023 & -0.024 & \\
R$_{\rm planet}$ / R$_{\star}$ & 0.0996 & +0.00055 & -0.00053 & \\
i [deg] & 87.44 & +0.18 & -0.20 & \\
a/R$_{\star}$ & 12.8 & +1.2 & -1.5 & \\
\hline
M$_{\star}$ [M$_{\odot}$] & 1.018 & +0.044 & -0.052 & (isochrone fit)\\
R$_{\star}$ [R$_{\odot}$] & 0.992 & +0.058 & -0.070 & (isochrone fit)\\
Age [Gyr] & 3.7 & +1.5 & -3.6 & (isochrone fit)\\
T$_{\rm eff}$ [K] & 5515 & +122 & -130 & (isochrone fit)\\
T$_{\rm eff}$ [K] & 5595 & +120 & -120 & (spectroscopic fit)\\
log L/L$_{\odot}$ & -0.087 & +0.078 & -0.088 & (isochrone fit)\\
log g [cgs] & 4.46 & +0.053 & -0.050 & (isochrone fit)\\
log g [cgs] & 4.23 & +0.2 & -0.2 & (spectroscopic fit)\\
$\rm {[Fe/H]}$ & 0.36 & +0.07 & -0.07 & (spectroscopic fit)\\
V$_{\rm rot}$ [km\,s$^{-1}$] & 2.0 & +2.0 & -2.0 & (spectroscopic fit)\\
\hline
R$_{\rm planet}$ [$R_{\oplus}$] & 10.8 & +0.63 & -0.77 & \\
R$_{\rm planet}$ [$R_{\rm Jup}$] & 0.96 & +0.06 & -0.07 & \\
K [m\,s$^{-1}$] & 78.7 & +8.5 & -9.5 & \\
$e \sin \omega$ & -0.123 & +0.089 & -0.110 & \\
$e \cos \omega$ & 0.053 & +0.086 & -0.079 & \\
M$_{\rm planet}$ [$M_{\oplus}$] & 209 & +24 & -28 & \\
M$_{\rm planet}$ [$M_{\rm Jup}$] & 0.66 & +0.08 & -0.09 & \\
a [AU] & 0.05714 & +0.00086 & -0.00093 & \\
$\rho_{\rm planet}$ [g\,cm$^{-3}$] & 0.93 & +0.18 & -0.22 & \\
\hline
\enddata
\end{deluxetable}


\begin{thebibliography}{}

\bibitem[2011]{nat11} Batalha, N., Borucki, W., Bryson, S., et al. 2011, \apj, 729, 27
\bibitem[2010]{tim10} Brown, T.~M. 2010, \apj, 709, 535
\bibitem[2010]{lars10} Buchhave, L.~A., Bakos, G.~A., Hartman, J.~D. et al.\,2010, \apj, 720, 1118
\bibitem[2010]{billb10} Borucki, W.~J., et al. 2010, Science, 327, 977
\bibitem[2011]{billb11} Borucki, W.~J., Koch, D.~J., Basri, G. et al. 2011, \apj, in press
\bibitem[2011]{erik11} Brugamyer, E., Dodson-Robinson, S. E., Cochran, W. D., Sneden, C. 2011, ApJ, submitted
\bibitem[2007]{adam07} Burrows, A., Hubeny, I., Budaj, J., Hubbard, W.~B. 2007, \apj, 661, 502
\bibitem[Djupvik \& Andersen(2010)]{djupvik:2010} Djupvik, A.~A., \&
  Andersen, J.\ 2010, in ``Highlights of Spanish Astrophysics V''
  eds. J.~M.~Diego, L.~J.~Goicoechea, J.~I.~Gonz\'alez-Serrano, \&
  J.~Gorgas (Springer: Berlin), p. 211
\bibitem[2011]{jmdes11} D\'esert, J.-M., Charbonneau, D., Demory, B.-O. et al., \apj, in prep.
\bibitem[2010]{ted10} Dunham, E.~W., Borucki, W.~J., Koch, D.~G., Batalha, N.~M., Buchhave, L.~A., et al. 2010, \apj, 713, L136
\bibitem[2000]{endl} Endl, M., K\"urster, M., \& Els, S. 2000, \aap, 362, 585
\bibitem[2007]{jona07} Fortney, J.~J., Marley, M.~S., \& Barnes, J.~W. 2007, \apj, 659, 1661
\bibitem[2011]{greg11} Gregory, P.~C. 2011, MNRAS, 410, 94
\bibitem[2006]{trist06} Guillot, T., Santos, N.~C., Pont, F. et al. 2006, \aap, 453, L21
\bibitem[2010]{horch} Horch, E. P. et al., 2010, AJ, 141, 45
\bibitem[2011]{howell} Howell, S. B., et al., 2011, AJ, 142, 19
\bibitem[2010]{jenk} Jenkins, J.~M., Caldwell, D.~A., Chandrasekaran, H., et al. 2010a, \apj, 713, L87
\bibitem[2010]{jenk2} Jenkins, J.~M., Borucki, W.~J., Koch, D.~G., et al. 2010b, \apj, 724, 1108 
\bibitem[1997]{mak} K\"urster, M., Schmitt, J. H. M. M., Cutispoto, G., \& Dennerl, K. 1997, \aap, 320, 831
\bibitem[1993]{kuru93} Kurucz, R. 1993, ATLAS9 Stellar Atmosphere Programs and 2 km/s grid. Kurucz CD-ROM No. 13. (Cambridge: Smithsonian Astrophys. Obs.)
\bibitem[latham]{dave10} Latham, D.~W., Borucki, W.~J., Koch, D.~G., et al. 2010, \apj, 713, L140
\bibitem[1976]{lom} Lomb, N.~R. 1976, Ap\&SS, 39, 447
\bibitem[2002]{mandel} Mandel, K. \& Algol, E. 2002, \apj, 580, L171
\bibitem[2009]{mill09} Miller, N., Fortney, J.~J., \& Jackson, B. 2009, \apj, 702, 1413
\bibitem[2011]{mill11} Miller, N., \& Fortney, J.~J. 2011, \apj, accepted
\bibitem[scargle]{scargl82} Scargle, J.~D. 1982, \apj, 263, 835
\bibitem[sneden]{chris73} Sneden, C. A. 1973, Ph.D. thesis, Univ. of Texas at Austin
\bibitem[sozetti]{soz2007} Sozetti, A., Torres, G., Charbonneau, D., Latham, D., Holman, M.~J., Winn, J.~N., Laird, J.~B., O'Donovan, F.~T. 2007, \apj, 664, 1190
\bibitem[ramsey]{larry} Ramsey, L.~W., et al. 1998, Proc. SPIE, 3352, 34
\bibitem[robin]{rob03} Robin, A.~C., Reyl\'e, C., Derri\'ere, S., Picaud, S. 2003, \aap, 409, 523
\bibitem[Torres et al.(2011)]{torres11} Torres, G. et al. 2011, \apj, 727, 24
\bibitem[tull]{tull95} Tull, R.~G., MacQueen, P.~J., Sneden, C., Lambert, D.~L. 1995, PASP, 107, 251
\bibitem[Tull]{tull98} Tull, R.~G. 1998, Proc. Soc. Photo-opt. Inst. Eng., 3355, 387
\bibitem[Yi]{yi2001} Yi, S., Demarque, P., Kim, Y.-C., Lee, Y.-W., Ree, C.~H., Lejeune, T., \& Barnes, S. 2001, ApJS, 136, 417

\end{thebibliography}
\end{document}